\documentclass[12pt]{article}
\usepackage{indent}
\usepackage{eqsection}
\usepackage{subeqnarray}
\usepackage{amsfonts}

\begin{document}

\bibliographystyle{plain}

\date{September 5, 2000 \\ Revised January 18, 2001}

\title{\vspace*{-1cm} Exact Finite-Size-Scaling Corrections to the \\
       Critical Two-Dimensional Ising Model 
       on a Torus}

\author{
  {\small Jes\'us Salas}                                    \\[-0.2cm]
  {\small\it Departamento de F\'{\i}sica Te\'orica}         \\[-0.2cm]
  {\small\it Facultad de Ciencias, Universidad de Zaragoza} \\[-0.2cm]
  {\small\it Zaragoza 50009, SPAIN}                         \\[-0.2cm]
  {\small\tt JESUS@MELKWEG.UNIZAR.ES}                       \\[-0.2cm]
  {\protect\makebox[5in]{\quad}}  % To force authors' names to be written
                                  %   vertically, one above another.
                                  % (\author seems to put them side-by-side
                                  %   if there is room.)
  \\
}
\vspace{0.5cm}

\maketitle
\thispagestyle{empty}   % Suppress page number on front page.

%%%%%%%%%%%%%%%%%%%%%%%%%%%%%%%%%%%%%%%%%%%%%%%%%%%%%%%%%%%%%%%%%%%%%%%%
%
% Begin definitions
%

%\ltapprox and \gtapprox produce > and < signs with twiddle underneath
\def\spose#1{\hbox to 0pt{#1\hss}}
\def\ltapprox{\mathrel{\spose{\lower 3pt\hbox{$\mathchar"218$}}
 \raise 2.0pt\hbox{$\mathchar"13C$}}}
\def\gtapprox{\mathrel{\spose{\lower 3pt\hbox{$\mathchar"218$}}
 \raise 2.0pt\hbox{$\mathchar"13E$}}}
\def\inapprox{\mathrel{\spose{\lower 3pt\hbox{$\mathchar"218$}}
 \raise 2.0pt\hbox{$\mathchar"232$}}}

%
% End definitions
%
%%%%%%%%%%%%%%%%%%%%%%%%%%%%%%%%%%%%%%%%%%%%%%%%%%%%%%%%%%%%%%%%%%%%%%%%

%%\doublespace

\begin{abstract}
We analyze the finite-size corrections to the energy and specific heat 
of the critical two-dimensional spin-1/2 Ising model on a torus.
We extend the analysis of Ferdinand and Fisher to compute the correction 
of order $L^{-3}$ to the energy and the corrections of order $L^{-2}$ and 
$L^{-3}$ to the specific heat. 
We also obtain general results on the 
form of the finite-size corrections to these quantities:
only integer powers of $L^{-1}$ occur, unmodified by logarithms 
(except of course for the leading $\log L$ term in the specific heat);
and the energy expansion contains only odd powers of $L^{-1}$. 
In the specific-heat expansion any power of $L^{-1}$ can appear, but the 
coefficients of the odd powers are
proportional to the corresponding coefficients of the energy expansion. 
\end{abstract}



\clearpage

%%%%%%%%%%%%%%%%%%%%%%%%%%%%%%%%%%%%%%%%%%%%%%%%%%%%%%%%%%%%%%%%%%%%%%%%
%
% Begin definitions 
%
\newcommand{\be}{\begin{equation}}
\newcommand{\ee}{\end{equation}}
\newcommand{\<}{\langle}
\renewcommand{\>}{\rangle}
\newcommand{\para}{\|}
\renewcommand{\perp}{\bot}

\def\smfrac#1#2{{\textstyle\frac{#1}{#2}}}
\def\half{ {{1 \over 2 }}}
\def\smhalf{ {\smfrac{1}{2}} }
\def\scra{{\cal A}}
\def\scrc{{\cal C}}
\def\scrd{{\cal D}}
\def\scre{{\cal E}}
\def\scrf{{\cal F}}
\def\scrg{{\cal G}}
\def\scrh{{\cal H}}
\def\scrj{{\cal J}}
\def\scrk{{\cal K}}
\def\scrl{{\cal L}}
\def\scrm{{\cal M}}
\newcommand{\scrmvec}{\vec{\cal M}_V}
\def\scrmtens{{\stackrel{\leftrightarrow}{\cal M}_T}}
\def\scro{{\cal O}}
\def\scrp{{\cal P}}
\def\scrr{{\cal R}}
\def\scrs{{\cal S}}
\def\ttens{{\stackrel{\leftrightarrow}{T}}}
\def\scrv{{\cal V}}
\def\scrw{{\cal W}}
\def\scry{{\cal Y}}
\def\tauss{\tau_{int,\,\scrm^2}}
\def\taux{\tau_{int,\,{\cal M}^2}}
\newcommand{\taum}{\tau_{int,\,\vec{\cal M}}}
\def\taue{\tau_{int,\,{\cal E}}}
\newcommand{\imag}{\mathop{\rm Im}\nolimits}
\newcommand{\real}{\mathop{\rm Re}\nolimits}
\newcommand{\tr}{\mathop{\rm tr}\nolimits}
\newcommand{\sgn}{\mathop{\rm sgn}\nolimits}
\newcommand{\codim}{\mathop{\rm codim}\nolimits}
\newcommand{\rank}{\mathop{\rm rank}\nolimits}
\newcommand{\sech}{\mathop{\rm sech}\nolimits}
\def\textprime{{${}^\prime$}}
\newcommand{\longto}{\longrightarrow}
\def\var{ \hbox{var} }
\newcommand{\gtilde}{ {\widetilde{G}} }
\newcommand{\USp}{ \hbox{\it USp} }
\newcommand{\CP}{ \hbox{\it CP\/} }
\newcommand{\QP}{ \hbox{\it QP\/} }
\def\hboxscript#1{ {\hbox{\scriptsize\em #1}} }

\newcommand{\plotdot}{\makebox(0,0){$\bullet$}}
\newcommand{\plotsmalldot}{\makebox(0,0){{\footnotesize $\bullet$}}}

\def\bsigma{\mbox{\protect\boldmath $\sigma$}}
\def\bpi{\mbox{\protect\boldmath $\pi$}}
\def\btau{\mbox{\protect\boldmath $\tau$}}
  % \boldmath is fragile, and without the \protect we get screwed when
  % we try to use \bsigma in a \caption.
\def\bn{{\bf n}}
\def\br{{\bf r}}
\def\bz{{\bf z}}
\def\bh{\mbox{\protect\boldmath $h$}}

\def\betatilde{ {\widetilde{\beta}} }
\def\hatp{\hat p}
\def\hatl{\hat l}

\def\msbar{ {\overline{\hbox{\scriptsize MS}}} }
\def\normalmsbar{ {\overline{\hbox{\normalsize MS}}} }

\def\eff{ {\hbox{\scriptsize\em eff}} }

\newcommand{\reff}[1]{(\ref{#1})}

\def\N{\hbox{$\mathbb N$}}
\def\C{\hbox{$\mathbb C$}}
\def\Q{\hbox{$\mathbb Q$}}
\def\R{\hbox{$\mathbb R$}}
\def\Z{\hbox{$\mathbb Z$}}

\newtheorem{theorem}{Theorem}[section]
\newtheorem{corollary}[theorem]{Corollary}
\newtheorem{lemma}[theorem]{Lemma}
\newtheorem{conjecture}[theorem]{Conjecture}
\newtheorem{definition}[theorem]{Definition}
\def\proof{\bigskip\par\noindent{\sc Proof.\ }}
\def\qed{\hbox{\hskip 6pt\vrule width6pt height7pt depth1pt \hskip1pt}\bigskip}

%
% Array for subscripts
%
\newenvironment{sarray}{
          \textfont0=\scriptfont0
          \scriptfont0=\scriptscriptfont0
          \textfont1=\scriptfont1
          \scriptfont1=\scriptscriptfont1
          \textfont2=\scriptfont2
          \scriptfont2=\scriptscriptfont2
          \textfont3=\scriptfont3
          \scriptfont3=\scriptscriptfont3
        \renewcommand{\arraystretch}{0.7}
        \begin{array}{l}}{\end{array}}

\newenvironment{scarray}{
          \textfont0=\scriptfont0
          \scriptfont0=\scriptscriptfont0
          \textfont1=\scriptfont1
          \scriptfont1=\scriptscriptfont1
          \textfont2=\scriptfont2
          \scriptfont2=\scriptscriptfont2
          \textfont3=\scriptfont3
          \scriptfont3=\scriptscriptfont3
        \renewcommand{\arraystretch}{0.7}
        \begin{array}{c}}{\end{array}}
%
% End definitions
%
%%%%%%%%%%%%%%%%%%%%%%%%%%%%%%%%%%%%%%%%%%%%%%%%%%%%%%%%%%%%%%%%%%%%%%%%
%
% Beginning of the text 
%
%%%%%%%%%%%%%%%%%%%%%%%%%%%%%%%%%%%%%%%%%%%%%%%%%%%%%%%%%%%%%%%%%%%%%%%%

%
% SECTION 1
%
\section{Introduction}   \label{sec_intro}

It is well-known that phase transitions in statistical-mechanical systems 
can occur only in the infinite-volume limit.  In any finite system,
all thermodynamic quantities (such as the magnetic susceptibility
and the specific heat) are analytic functions of all parameters
(such as the temperature and the magnetic field);
but near a critical point they display peaks whose height increases
and whose width decreases as the volume $N=L^d$ grows,
yielding the critical singularities in the limit $L \to\infty$.
For bulk experimental systems (containing $N \sim 10^{23}$ particles)
the finite-size rounding of the phase transition is usually beyond
the experimental resolution;
but in Monte Carlo simulations ($N\ltapprox 10^6$--$10^7$)
it is visible and is often the dominant effect.

Finite-size scaling theory
\cite{Fisher,Fisher_Barber,Barber,Privman}
provides a systematic framework for understanding finite-size effects
near a critical point.
The idea is simple:  the only two relevant length scales are
the system linear size $L$ and the correlation length $\xi_\infty$ of the
bulk system at the same parameters, so everything is controlled by the
single ratio $\xi_\infty/L$.\footnote{
   This is true only for systems below the upper critical dimension $d_c$. 
   For Ising models with short-range interaction, $d_c = 4$.
}
If $L \gg \xi_\infty$, then finite-size effects are negligible;
for $L \sim \xi_\infty$, thermodynamic singularities are rounded
and obey a scaling Ansatz
${\cal O} \sim L^{p_{\cal O}} F_{\cal O}(\xi_\infty/L)$
where $p_{\cal O}$ is a critical exponent and
$F_{\cal O}$ is a scaling function.
Finite-size scaling is the basis of the powerful 
phenomenological renormalization group method
(see Ref.~\cite{Barber} for a review);
and it is an efficient tool for extrapolating finite-size data
coming from Monte Carlo simulations so as to obtain accurate results on
critical exponents, universal amplitude ratios and subleading exponents
\cite[and references therein]{fss_greedy,MGMC_SU3,%
Salas_Sokal_Ising_published,XY}.\footnote{
  Finite-size scaling has also been successfully applied to data coming 
  from transfer-matrix computations \cite{Blote}.
} 
In particular, in systems with multiplicative and/or additive logarithmic
corrections (as the two-dimensional 4-state Potts model \cite{Salas_Sokal_FSS}),
a good understanding of finite-size effects is crucial for obtaining reliable
estimates of the physically interesting quantities.

In finite-size-scaling theory for systems with periodic boundary conditions,
three simplifying assumptions have frequently been made:
\begin{itemize}
\item[(a)] The regular part of the free energy, $f_{\rm reg}$,
           is independent of the lattice size $L$ \cite{Privman}
           (except possibly for terms that are exponentially small in $L$).
\item[(b)] The scaling fields associated to the temperature and magnetic 
           field (i.e., $g_t$ and $g_h$, respectively) are independent of $L$
           \cite{Guo_87}.
\item[(c)] The scaling field $g_L$ associated to the lattice size 
           equals $L^{-1}$ exactly, with no corrections $L^{-2}$,
           $L^{-3}$, \ldots\  \cite{Privman}.
\end{itemize}
Moreover, in the nearest-neighbor spin-1/2 2D Ising model,
it has further been assumed that
\begin{itemize}
\item[(d)]  There are no irrelevant operators \cite{Aharony_83,Gartenhaus_88}.
\end{itemize}
Unfortunately, the combination of these four assumptions implies
that the asymptotic expansions for the energy and specific heat
for the Ising model at criticality terminate at order $1/L$ 
(see Ref.~\cite{Salas_Sokal_Ising}). 
However, the numerical results presented in 
\cite{Salas_Sokal_Ising,Salas_Sokal_Ising_published}  
(as well as the analytic results presented in this paper) show this 
to be false. The problem, therefore, is to determine which one(s) of these 
assumptions are invalid, and why.
Assumption (c) is extremely plausible from renormalization-group  
considerations,
at least for periodic boundary conditions; 
and assumption (d) has been confirmed numerically through order $(T-T_c)^3$
at least as regards the bulk behavior of the susceptibility
\cite{Gartenhaus_88}. However, both numerical \cite{Nickel_a,Nickel_b}
and theoretical \cite{Pelissetto_99} evidence
has recently emerged suggesting that irrelevant operators do contribute 
to the susceptibility at order $(T-T_c)^4$.

In a classic paper, Ferdinand and Fisher \cite{Ferdinand_Fisher} 
considered the energy and the specific heat of the two-dimensional Ising model
on a torus of length $L$ and aspect ratio $\rho$,
and obtained the first two (resp. three) terms
of the large-$L$ asymptotic expansion of the energy (resp.\ specific heat)
at fixed $x \equiv L(T-T_c)$ [this is the finite-size-scaling regime]
and fixed $\rho$.\footnote{
  Ferdinand and Fisher \protect\cite{Ferdinand_Fisher} also obtained the 
  position of the maximum of the specific heat $x_{\rm max}(\rho)$ which 
  depends on the torus aspect ratio $\rho$. Furthermore, Kleban and Akinci 
  \protect\cite{Kleban1,Kleban2} showed that an excellent approximation 
  can be obtained by keeping only the two largest eigenvalues of 
  the transfer matrix, and they interpreted their results in terms of 
  domain-wall energies. This approximation is already good at $\rho=1$ and  
  becomes exponentially better with increasing $\rho$. 
  This method could surely simplify the computations 
  presented in this paper; but here we are interested in the {\em exact} 
  values of the finite-size-scaling corrections. These results are used as
  theoretical inputs in \protect\cite{Salas_Sokal_Ising,Salas_Sokal_in_prep}. 
} 
In particular, at criticality ($T=T_c$) 
they computed the finite-size corrections to both quantities to order $L^{-1}$. 
In 1999, Hu {\em et al}.\ \cite{Izmailian} published (without details) 
the correction of order $L^{-3}$ to the energy and showed that the 
$L^{-2}$ correction is absent.\footnote{
   The published version of \protect\cite{Izmailian} contains
   several misprints in the crucial formula \protect\reff{def_E3}.  
   Version 2 of this paper in the Los Alamos preprint archive cond-mat
   contains the correct formula.}
In this paper we compute explicitly the following finite-size corrections: 

\begin{itemize}

\item The correction of order $L^{-3}$ to the energy. 

\item The corrections of order $L^{-2}$ and $L^{-3}$ to the specific heat. 

\end{itemize}
Furthermore, we also find new insights into the general 
analytic structure of the finite-size corrections to this model. 
We show that in the critical two-dimensional Ising model:

\begin{itemize}

\item The finite-size corrections to the energy and specific heat
are always integer powers of $L^{-1}$, {\em unmodified by logarithms}\/
(except of course for the leading $\log L$ term in the specific heat).

\item In the finite-size expansion of the energy, only {\em odd}\/ integer 
powers of $L^{-1}$ occur.  

\item In the finite-size expansion of the specific heat, any integer powers
of $L^{-1}$ can occur. However, the coefficients of the odd powers 
of $L^{-1}$ in this expansion are proportional to the corresponding coefficients
in the energy expansion.  

\end{itemize}
These results can be compared to the general renormalization-group
expression for the finite-size 
corrections to the energy and the specific heat \cite{Salas_Sokal_Ising},
in which arbitrary powers of $L^{-1}$ and  
terms of the type $L^{-1+p} \log L$ (where $p$ is some real number)
can occur.
The implications of these results for understanding which one(s) of the 
assumptions (a)--(d) are invalid will be analyzed 
elsewhere \cite{Salas_Sokal_in_prep}. 

The plan of this paper is as follows:
In Section~\ref{sec_definitions} we present our definitions and notation 
(generally following Ref.~\cite{Ferdinand_Fisher}).
In Sections~\ref{sec_energy} and
\ref{sec_specific_heat} we present the computation of the next terms in the
asymptotic expansions for the energy and specific heat, respectively.
Finally, in Section~\ref{sec_conclusions} we present our arguments 
about the type of finite-size-scaling corrections that can occur in these two
expansions. We have summarized in Appendix~\ref{sec_theta} the basic 
definitions and properties of the $\theta$-functions that will be needed in
this paper. In Appendix~\ref{sec_Euler} we recall the Euler-MacLaurin formula. 

\bigskip 

\noindent
{\bf Remark}. After the completion of this work, we learned that similar 
results have been independently obtained by Izmailian and Hu \cite{Izmailian2}.

%
% SECTION 2
%
\section{Basic definitions} \label{sec_definitions} 

Let us consider an Ising model on a torus of size $m \times n$ at zero
magnetic field. 
The Hamiltonian is given by\footnote{In this paper we are following 
basically the notation used by Ferdinand and Fisher in 
\protect\cite{Ferdinand_Fisher}, with a few minor modifications.
} 
\be
{\cal H} = - K \sum\limits_{\<i,j\>} \sigma_i \sigma_j 
\label{def_hamiltonian}
\ee
The partition function can be written as 
\be
Z_{mn} = \sum\limits_{\{\sigma\}} e^{-{\cal H}} = 
      {1 \over 2} (2 \sinh 2K)^{mn/2} \sum\limits_{i=1}^4 Z_i(K,n,m) 
 \;,
\label{def_partition_function}
\ee
where the partial partition function $Z_1$ is given by 
\be
Z_1(K,n,m) = \prod\limits_{r=0}^{n-1} 2 \cosh \left( {m \gamma_{2r+1} \over 2}
           \right)
\label{def_z1}
\ee
and the rest are defined analogously using the first three columns of the 
following table [the last two columns will be needed afterwards in 
\reff{def_z_ratios}/\reff{Q14_final_all}].  
\be
\begin{array}{lcrrr}
Z_1:\quad  & 2r + 1 & \cosh & \tanh & \sech          \\
Z_2:       & 2r + 1 & \sinh & \coth & i\, {\rm csch} \\
Z_3:       & 2r     & \cosh & \tanh & \sech          \\
Z_4:       & 2r     & \sinh & \coth & i\, {\rm csch} \\
\end{array}
\label{table_notation}
\ee
The quantities $\gamma_l = \gamma_l(K,n)$ are defined by 
\be
\cosh \gamma_l \equiv c_l = 
      \cosh 2K \coth 2K - \cos\left( {l \pi \over n}\right)
\label{def_cosh_gamma_l}
\ee
In particular, we have 
\begin{subeqnarray}
\slabel{del_gamma_0}
\gamma_0 &=& 2K + \log (\tanh K) \\
\gamma_l &=& \log \left( c_l + \sqrt{c_l^2 - 1} \right) \qquad l\neq 0
\slabel{def_gamma_l}
\label{def_gammas}
\end{subeqnarray}
The quantities $\gamma_l$ \reff{def_gammas} satisfy $\gamma_l = \gamma_{2n-l}$,
and $\gamma_l$ is a monotonically increasing function of $l$ for 
$0\leq l \leq n$.

The internal energy density $E$ and the specific heat $C_H$ are given by
\begin{eqnarray}
       \label{def_energy}
E(K,m,n) &=& - \coth 2K - {1 \over mn} \left[  
       { \sum\limits_{i=1}^4 Z_i^\prime \over \sum\limits_{i=1}^4 Z_i} 
       \right] \\
C_H(K,m,n) &=& - 2{\rm csch}^2 2K + {1 \over mn}
     \left[ 
     { \sum\limits_{i=1}^4 Z_i^{\prime\prime} \over \sum\limits_{i=1}^4 Z_i}
     - \left(
     { \sum\limits_{i=1}^4 Z_i^\prime \over \sum\limits_{i=1}^4 Z_i} 
     \right)^2 \right]
\label{def_specific_heat}
\end{eqnarray}
where the primes denote derivatives with respect to the coupling constant
$K$. 
In computing observables \reff{def_energy}/\reff{def_specific_heat},
the following formulae, derived from \reff{def_z1}, will be useful:
\begin{subeqnarray}
\slabel{def_z1prime_over_z1}
{Z_1^\prime \over Z_1} &=& {m \over 2} \sum\limits_{r=0}^{n-1} 
     \gamma^\prime_{2r+1} \tanh\left( {m \gamma_{2r+1} \over 2} \right) \\
{Z_1^{\prime\prime} \over Z_1} &=& 
     \left[ {m \over 2} \sum\limits_{r=0}^{n-1} \gamma^\prime_{2r+1} 
            \tanh\left({m \gamma_{2r+1} \over 2}\right) \right]^2 + 
     {m \over 2} \sum\limits_{r=0}^{n-1} \gamma^{\prime\prime}_{2r+1} 
            \tanh\left({m \gamma_{2r+1} \over 2}\right) \nonumber \\
     & & \qquad + \left({m \over 2}\right)^2 \sum\limits_{r=0}^{n-1}
         \left[ \gamma^\prime_{2r+1} 
                \sech\left({m \gamma_{2r+1} \over 2}\right) \right]^2 
\slabel{def_z1doubleprime_over_z1}
\label{def_z_ratios}
\end{subeqnarray}
The analogous ratios for $i=2,3,4$ can be obtained from 
\reff{def_z1prime_over_z1}/\reff{def_z1doubleprime_over_z1} by using 
\reff{table_notation} [Note that the third column of \reff{table_notation} 
does not play any role here].
The factor $i$ in the entry $i\, {\rm csch}$ of \reff{table_notation} 
changes the sign of the last term of 
\reff{def_z1doubleprime_over_z1} for $Z_2^{\prime\prime}$ and 
$Z_4^{\prime\prime}$.

The critical point of the Ising model corresponds to the self-dual point 
$\sinh 2K_c=1$. That is 
\be
K_c = {1 \over 2} \log(1 + \sqrt{2})
\label{def_critical_point}
\ee
In this paper we are concerned with the finite-size-scaling corrections 
to the energy and specific heat of the critical Ising model. 
We will express all our results in terms of the length $n$ and the aspect 
ratio of the torus $\rho$:\footnote{In conformal-field-theory language, the 
modular parameter of the torus is $\tau = i\,\rho$ \protect\cite{Itzykson}, 
where this $\tau$ has nothing to do with the temperature-like parameter 
defined in \protect\reff{def_tau_over_n}.
}
\be
\rho = {m \over n}
\label{def_aspect_ratio}
\ee
Indeed, all our results are invariant under the transformation 
$n \leftrightarrow m$. Here we shall show that the energy and specific 
heat at criticality have asymptotic expansions of the form
\begin{eqnarray}
\label{expansion_energy}
-E(K_c,m,n) &\equiv& -E_c(n,\rho) = E_0 + \sum\limits_{k=1}^\infty 
           {E_k(\rho) \over n^k} \\ 
C_H(K_c,n,m) &\equiv& C_{H,c}(n,\rho) = C_{00} \log n + C_0(\rho) + 
  \sum\limits_{k=1}^\infty {C_k(\rho) \over n^k} 
\label{expansion_specific_heat}
\end{eqnarray}
The coefficients $E_0$ and $C_{00}$ can be obtained from Onsager's solution
\cite{Onsager};
$E_1$, $C_0$, and $C_1$ were computed by Ferdinand and 
Fisher \cite{Ferdinand_Fisher}; and finally,
the fact that $E_2 = 0$ and the expression for $E_3$ were given 
(without details) in Ref.~\cite{Izmailian}. Here we shall compute explicitly
the terms $E_3$, $C_2$ and $C_3$, and shall show that
$E_2 = E_4 = E_6 = \ldots = 0$.

Let us now see how $\gamma_l$ and its derivatives behave close
to the critical point \reff{def_critical_point}. To do so, we introduce the 
finite-size-scaling
parameter $\tau$ as in Ref.~\cite[Eq.~(2.12)]{Ferdinand_Fisher}\footnote{
  The parameter $\tau$ plays the same role as the usual finite-size-scaling
  parameter $x \equiv L(T-T_c)$.
  Indeed, to leading order in $K-K_c$ we have $\tau = - 2 n (K-K_c)$.
}:
\be
\left( {\tau \over n} \right)^2 = {1 \over 2}\left( \sinh 2K + 
       {1 \over \sinh 2K} \right) - 1
\label{def_tau_over_n}
\ee
Thus, $\tau=0$ corresponds to the critical point $K=K_c$,
and $\tau \neq 0$ fixed corresponds to the finite-size-scaling regime
$n \to\infty$, $K \to K_c$ with $n(K-K_c)$ fixed.
 Hereafter, we will 
consider the behavior of all quantities as a function of $\tau$ in the 
limit $\tau\rightarrow 0$. 
The value of $\gamma_0$ at $\tau=0$ is zero; its behavior close to the 
critical point is given by 
\be
\label{gamma_0_T=Tc}
\gamma_0(\tau,n) = - 2 \left( {\tau \over n} \right) + 
                     {\cal O}\left[ \left({\tau \over n}\right)^3 \right] 
\ee
The derivatives of $\gamma_0$ with respect to $K$ are non-vanishing at
criticality:  
\begin{subeqnarray}
\slabel{gamma_prime_0_T=Tc}
\gamma_0^\prime(0,n) &\equiv& \left.{d\gamma_0 \over dK}\right|_{T=T_c} = 4 \\
\slabel{gamma_doubleprime_0_T=Tc}
\gamma_0^{\prime\prime}(0,n) &\equiv& 
          \left.{d^2\gamma_0 \over dK^2}\right|_{T=T_c} = - 4 \sqrt{2} 
\label{gamma_0_derivatives_T=Tc}
\end{subeqnarray}
(Note that prime continues to denote $d/dK$, {\em not} $d/d\tau$. However, 
the final result will be expressed in terms of $\tau$ [in the limit 
$\tau\rightarrow 0$], hence the notation $\gamma_0^\prime(0,n)$). 
For a generic $l \neq 0$ the critical value of $\gamma_l$ is given by
\be
\label{gamma_l_T=Tc}
\gamma_l(0,n) =
       2 \log\left[ \sqrt{1 + \sin^2\left({l \pi \over 2 n}\right)} +
             \sin\left({l \pi \over 2 n}\right) \right]
\ee
while its derivatives with respect to $K$ are given by
\begin{subeqnarray}
\slabel{gamma_prime_l_T=Tc}
\gamma_l^\prime &=& {c_l^\prime \over \sqrt{c_l^2 - 1}} \\
\gamma_l^{\prime\prime} &=& {c_l^{\prime\prime} \over \sqrt{c_l^2 - 1}} - 
             {c_l (c_l^{\prime})^2\over (c_l^2 - 1)^{3/2}} 
\slabel{c_doubleprime_l_T=Tc}
\end{subeqnarray}
where the quantity $c_l^\prime$ vanishes at criticality as 
\be
\label{c_prime_l_T=Tc}
c_l^\prime(\tau,n)  = c^\prime(\tau,n) = - 8 \left( {\tau \over n} \right) 
+ {\cal O}\left[ \left( {\tau \over n} \right)^2 \right] 
\ee
and the quantity $c_l^{\prime\prime}$ gives a non-zero value
\be
\label{gamma_doubleprime_l_T=Tc}
c_l^{\prime\prime}(0,n) = c^{\prime\prime}(0,n) = 16 
\ee 

We can write the partial partition functions $Z_i$ 
[cf.\ \reff{def_z1}/\reff{table_notation}] in the following form:
\begin{subeqnarray}
Z_1(\tau,n,\rho) &=& P_1(\tau,n,\rho) 
                    \exp\left( {m \over 2} \sum\limits_{r=0}^{n-1} 
                    \gamma_{2r+1} \right) \\
Z_2(\tau,n,\rho) &=& P_2(\tau,n,\rho) 
                    \exp\left( {m \over 2} \sum\limits_{r=0}^{n-1} 
                    \gamma_{2r+1} \right) \\
Z_3(\tau,n,\rho) &=& P_3(\tau,n,\rho) 
                    \exp\left( {m \over 2} \sum\limits_{r=0}^{n-1} 
                    \gamma_{2r} \right) \left[1 + e^{-m\gamma_0}\right] \\
Z_4(\tau,n,\rho) &=& P_4(\tau,n,\rho) 
                    \exp\left( {m \over 2} \sum\limits_{r=0}^{n-1} 
                    \gamma_{2r} \right) \left[1 - e^{-m\gamma_0}\right] 
\label{def_partial_partition_functions}
\end{subeqnarray}
where the quantities $P_i(\tau,n,\rho)$ are given by 
\begin{subeqnarray}
\log P_1(\tau,n,\rho) &=& 
         \sum\limits_{r=0}^{n-1} \log( 1 + e^{-m\gamma_{2r+1}} )\\
\log P_2(\tau,n,\rho) &=& 
         \sum\limits_{r=0}^{n-1} \log( 1 - e^{-m\gamma_{2r+1}} )\\
\log P_3(\tau,n,\rho) &=& 
         \sum\limits_{r=1}^{n-1} \log( 1 + e^{-m\gamma_{2r}} )  \\
\log P_4(\tau,n,\rho) &=& 
         \sum\limits_{r=1}^{n-1} \log( 1 - e^{-m\gamma_{2r}} ) 
\label{def_P}
\end{subeqnarray}
The functions \reff{def_P} give non-vanishing constants in the limit 
$\tau\rightarrow 0$ \cite{Ferdinand_Fisher}:
\begin{subeqnarray}
\log P_1(0,n,\rho) &=& {\theta_3 \over \theta_0} + 
                      {\cal O}(n^{-2}) \\
\log P_2(0,n,\rho) &=& {\theta_4 \over \theta_0} + 
                      {\cal O}(n^{-2}) \\
\log P_3(0,n,\rho) &=& {1 \over 2} {\theta_2 \over \theta_0} e^{\pi\rho/4} +
                      {\cal O}(n^{-2}) \\
\log P_4(0,n,\rho) &=& \theta_0^2 + 
                      {\cal O}(n^{-2}) 
\slabel{def_P4_T=Tc}
\label{def_P_T=Tc}
\end{subeqnarray}
where the functions $\theta_i$ with $i=2,3,4$ are the usual $\theta$-functions 
(see Appendix~\ref{sec_theta}), and $\theta_0$ is defined in 
\reff{def_theta0}. 

Finally, we introduce the ratios 
\be 
R_i(\tau,n,\rho) = {Z_i(\tau,n,\rho) \over Z_1(\tau,n,\rho)}
\label{def_ratios_R}
\ee
Thus, from \reff{def_partial_partition_functions} we get 
\begin{subeqnarray}
R_1(\tau,n,\rho) &=& 1 \\
R_2(\tau,n,\rho) &=& {P_2(\tau,n,\rho) \over P_1(\tau,n,\rho)} \\
R_3(\tau,n,\rho) &=& 2 \cosh\left( {m \gamma_0 \over 2} \right) 
                    P_0(\tau,n,\rho) {P_3(\tau,n,\rho) \over P_1(\tau,n,\rho)} \\
R_4(\tau,n,\rho) &=& 2 \sinh\left( {m \gamma_0 \over 2} \right) 
                    P_0(\tau,n,\rho) {P_4(\tau,n,\rho) \over P_1(\tau,n,\rho)} 
\label{ratios_R}
\end{subeqnarray}
where $P_0(\tau,n,\rho)$ is defined as
\be
\log P_0(\tau,n,\rho) = {m \over 2} \left[ \sum\limits_{r=1}^{n-1} 
              \gamma_{2r} - \sum\limits_{r=0}^{n-1} \gamma_{2r+1} \right]
\label{def_P0}
\ee
The sum of the four ratios is denoted by $R$
\be
R(\tau,n,\rho) = \sum\limits_{i=1}^4 R_i(\tau,n,\rho)
\label{def_R}
\ee

The ratios $R_2$ and $R_3$ have a non-vanishing value at the critical point
\cite{Ferdinand_Fisher}
\begin{subeqnarray} 
R_2(0,n,\rho) &=& {\theta_4 \over \theta_3} + 
                 {\cal O}(n^{-2}) \\
R_3(0,n,\rho) &=& {\theta_2 \over \theta_3} + 
                 {\cal O}(n^{-2})  
\label{ratios_R2_R3_T=Tc}
\end{subeqnarray} 
while $R_4$ vanishes at $K=K_c$:
\be
R_4(\tau,n,\rho) = - \sinh(\tau\rho) \left[ \theta_2 \theta_4 + 
{\cal O}(n^{-2}) \right] + {\cal O}(\tau^2)  
\label{ratio_R4_T=Tc}
\ee
The sum of the four ratios at criticality is a non-zero constant 
\be
R(0,n,\rho) = {\theta_2 + \theta_3 + \theta_4 \over \theta_3} + {\cal O}(n^{-2}) 
\label{R_T=Tc}
\ee
The function $P_0$ has also a non-vanishing limit at criticality:
\be
P_0(0,n,\rho) = e^{-\pi\rho/4}\left[1 + {\cal O}(n^{-2})\right] 
\label{P0_T=Tc}
\ee
%

%
% SECTION 3
%
\section{Finite-size-scaling corrections to the internal energy}
\label{sec_energy}

The internal energy at the critical point $E_c$ is equal to 
\be
-E_c(n,\rho)  = 
-E(K_c,n,\rho) = \sqrt{2} + \lim\limits_{\tau\rightarrow0} 
                {1 \over m n R} \, \sum\limits_{i=1}^4 
                {Z_i^\prime \over Z_i} \, R_i  
\label{energy_T=Tc}
\ee
The terms $Z_i^\prime/Z_i$ with $i=1,2$ vanish trivially as all 
the $\gamma^\prime_{2r+1}$ vanish. The term $Z_3^\prime/Z_3$ does not vanish 
due to the contribution of $\gamma_0^\prime$; but its total contribution is
also zero as it is multiplied by $\tanh (m\gamma_0/2)$, which vanishes at
criticality. The only non-vanishing contribution comes from $i=4$:  
\be
{Z_4^\prime \over Z_4} = {m \over 2} \gamma_0^\prime \coth
          \left({m \gamma_0 \over 2} \right)  \sim  
          - 2 m \coth (\rho \tau) 
           \quad {\rm as} \quad \tau \rightarrow 0
\label{z4_prime_over_z4_T=Tc}
\ee
So we obtain the formula 
\be
-E_c(n,\rho) = \sqrt{2} - {2 \over n} \lim\limits_{\tau \rightarrow 0} 
                {R_4(\tau,n,\rho) \over R(0,n,\rho)} \coth(\rho \tau)
\label{energy_T=Tc_final}
\ee
where $R(0,n,\rho)$ is given by \reff{R_T=Tc}.
Note that, by \reff{ratio_R4_T=Tc},
$R_4(\tau,n,\rho) \sim \tau$ as $\tau\rightarrow 0$,
so $R_4(\tau,n,\rho) \coth(\rho \tau)$ gives rise to a non-zero result in
this limit.

The goal of this section is to extend the Ferdinand-Fisher asymptotic expansion
\cite{Ferdinand_Fisher} to order $n^{-4}$. Let us first consider the 
quantity $\log P_4$ at criticality:
\be
\log P_4(0,n,\rho) = 
  \sum\limits_{r=1}^{n-1} \log\left( 1 - e^{-m\gamma_{2r}} \right)
= -2 \sum\limits_{p=1}^\infty {1\over p} 
     \sum\limits_{r=1}^{\lfloor {n\over 2} \rfloor} e^{-mp\gamma_{2r}}
\label{logP4_1}
\ee
where $\lfloor x \rfloor$ is the largest integer $\le x$.
The sum over $r$ can be split into two parts:
\be
\sum\limits_{r=1}^{\lfloor {n\over 2} \rfloor} = 
\sum\limits_{r=1}^{s(n)-1} + \sum\limits_{r=s(n)}^{\lfloor {n\over 2} \rfloor } 
\label{sum_splitting_trick}
\ee
where $s(n)$ will be chosen afterwards. We can drop the second sum in 
\reff{sum_splitting_trick}, as it gives a contribution of order 
$\sim n | \log[1 - \exp(-m\gamma_{2s}) ]| \sim n \exp[-2\pi \rho s(n)]$.  
Instead of the choice $s(n) = (3/2\pi\rho) \log n$
made by Ferdinand and Fisher, we shall use the choice
\be
s(n) = {M \over 2 \pi \rho} \log n 
\label{def_s}
\ee
with $M$ an arbitrary positive integer,
to ensure that the total contribution of the second sum in 
\reff{sum_splitting_trick} is as small as we want (namely, $\sim n^{-(M-1)}$). 

We can use the following expression for $\gamma_{2r}$ at criticality 
\cite{Ferdinand_Fisher}:
\be
{1 \over 2}\gamma_{2r}(0,n) = \log\left[ \sin\left({r \pi \over n}\right) + 
  \sqrt{ 1 + \sin^2\left({r \pi \over n}\right)} \right]
\label{expression_gamma_l} 
\ee
to obtain an asymptotic series of $m\gamma_{2r}$ in terms of $n^{-1}$:
\be
m \gamma_{2r}(0,n) = 2 \pi \rho r - {2 \rho \pi^3 \over 3} {r^3 \over n^2}  
 + {\rho \pi^5\over 3} {r^5 \over n^4} + {\cal O}(n^{-6})
\label{series_m_gamma_2r}
\ee
We should recall that the choice of $s(n)$ \reff{def_s} for {\em any} integer 
$M$ guarantees that the ratio $r/n$ is a small quantity for $0\leq r \leq s(n)$ 
(as $r/n \leq s(n)/n \ll 1$ if $n$ is large enough). 

\bigskip

\noindent
{\bf Remark:} 
It is interesting to note that only even powers of $n^{-1}$ occur in the 
expansion \reff{series_m_gamma_2r}.

\bigskip

Plugging the expansion \reff{series_m_gamma_2r} in Eq.~\reff{logP4_1} we 
obtain 
\be
\log P_4(0,n,\rho) = -2 \sum\limits_{p=1}^\infty {1 \over p} 
              \sum\limits_{r=1}^{s(n)-1} e^{-2\pi r p \rho} - 
          {4 \pi^3 \rho\over 3} \sum\limits_{p=1}^\infty
              \sum\limits_{r=1}^{s(n)-1} {r^3 \over n^2} 
            e^{-2\pi r p \rho} + {\cal O}(n^{-4})
\label{logP4_2}
\ee
We can extend the sums $\sum_{r=1}^{s(n)-1}$ to $\sum_{r=1}^\infty$ in 
\reff{logP4_2} by introducing an error of order 
$n^{-M}$. 
The second term of the r.h.s. of \reff{logP4_2} can be expressed in terms of 
\be
\sum\limits_{r=1}^\infty r^3 e^{-2\pi r p \rho} = 
  {1 \over 4} \left[ {1\over \sinh^2(\pi p \rho)} + {3\over 2}
                     {1\over \sinh^4(\pi p \rho)} \right]
\label{def_I4}
\ee
Finally, we can write $P_4(n,\rho)$ in the following form
\be
P_4(0,n,\rho) = \theta_0^2\left[ 1 - {1\over n^2}\, {p_1(\rho) \over 2} + 
            {\cal O}(n^{-4}) \right]
\label{P4_final}
\ee
which improves \reff{def_P4_T=Tc}. The function $p_1(\rho)$ is given by
\be
p_1(\rho) = {2\pi^3 \rho \over 3} \sum\limits_{m=1}^\infty
   \left[ {1 \over \sinh^2(m \pi \rho) } + {3\over 2}{1\over \sinh^4(m \pi \rho) }
   \right] \\
\label{def_p1}
\ee

Using similar methods one can obtain the improved version of \reff{def_P_T=Tc}:
\begin{subeqnarray}
P_1(0,n,\rho) &=& {\theta_3 \over \theta_0} \left[ 1 - 
     {1 \over n^2}\, \left(\widetilde{p}_2 - {\widetilde{p}_1 \over 2}
                     \right) + {\cal O}(n^{-4}) \right] \\
P_2(0,n,\rho) &=& {\theta_4 \over \theta_0} \left[ 1 +  
     {1 \over n^2}\, \left({p_1\over 2} - p_2 
                     \right) + {\cal O}(n^{-4}) \right] \\ 
P_3(0,n,\rho) &=& {1\over 2}{\theta_2 \over \theta_0} e^{\pi\rho/4} 
                      \left[ 1 +  
     {1 \over n^2}\, {\widetilde{p}_1\over 2}  
                     + {\cal O}(n^{-4}) \right]
\label{P's_final}
\end{subeqnarray}
where $p_2$, $\widetilde{p}_1$, and $\widetilde{p}_2$ are defined by 
\begin{subeqnarray}
\widetilde{p}_1(\rho) &=& {2\pi^3 \rho \over 3} \sum\limits_{m=1}^\infty
   (-1)^{m+1}
   \left[ {1 \over \sinh^2(m \pi \rho) } + {3\over 2}{1\over \sinh^4(m \pi \rho) }
   \right] \\
p_2(\rho) &=& {\pi^3 \rho \over 24} \sum\limits_{m=1}^\infty
   \left[ {1 \over \sinh^2(m \pi \rho/2) } +
   {3\over 2}{1\over \sinh^4(m \pi \rho/2) }
   \right] \\
\widetilde{p}_2(\rho) &=& {\pi^3 \rho \over 24} \sum\limits_{m=1}^\infty
   (-1)^{m+1}
   \left[ {1 \over \sinh^2(m \pi \rho/2) } +
   {3\over 2}{1\over \sinh^4(m \pi \rho/2) }
   \right] 
\label{def_p's}
\end{subeqnarray}

Finally, we have to improve the expression of $\log P_0$ \reff{P0_T=Tc}.
Let us first consider the sum
\be
{m\over 2}\sum\limits_{r=1}^{n-1}\gamma_{2r}(0,n) = m \sum\limits_{r=0}^{n-1}  
\log \left[ \sin\left({r \pi \over n}\right) + 
     \sqrt{ 1 + \sin^2\left({r \pi \over n}\right)} \right] 
\label{logP0_1_1}
\ee
We can apply the Euler-MacLaurin formula \reff{Euler_MacLaurin_formula}
to the function  $f(p) = \log(\sin p + \sqrt{1 + \sin^2 p})$ with 
$L=2n$ and $\alpha=1/2$. The result is\footnote{It is easy to 
verify that $f^{(3)}(p)$ is integrable over $[0,\pi]$. This means that the
next term in the expansion \protect\reff{logP0_1_final} is of order 
${\cal O}(n^{-4})$.}
\be
{m\over 2}\sum\limits_{r=1}^{n-1} \gamma_{2r}(0,n)= 
{2 m n \over \pi} G -{\pi\rho\over 6} - {\pi^3\rho\over 180} {1 \over n^2} + 
        {\cal O}(n^{-4})  
\label{logP0_1_final}
\ee
where $G \approx 0.915965594177219$ is Catalan's constant. Using similar 
methods we obtain the other sum appearing in \reff{P0_T=Tc}: \footnote{
  The leading terms of 
  Eqs.~\protect\reff{logP0_1_final}/\protect\reff{logP0_2_final} were obtained
  by Ferdinand \protect\cite{Ferdinand}.
}
\be
{m\over 2}\sum\limits_{r=0}^{n-1} \gamma_{2r+1}(0,n) =
{2 m n \over \pi} G -{\pi\rho\over 12} - {7\pi^3\rho\over 1440} {1 \over n^2} +
        {\cal O}(n^{-4})
\label{logP0_2_final}
\ee
Putting \reff{logP0_1_final} and \reff{logP0_2_final} together we obtain the 
improved version of \reff{P0_T=Tc}
\be
P_0(0,n,\rho) = e^{-\pi\rho/4}\left[ 1 - {p_3(\rho) \over n^2} 
           + {\cal O}(n^{-4}) \right]
\label{P0_final}
\ee
where $p_3(\rho)$ is defined as 
\be
p_3(\rho) = {\pi^3 \over 96} \rho 
\label{def_p3}
\ee

The improved expressions for the ratios $R_i(\tau=0,n,\rho)$ are easily 
obtained from \reff{P4_final}/\reff{P's_final} 
\begin{subeqnarray}
R_2(0,n,\rho) &=& {\theta_4 \over \theta_3} \left[ 1 - {1\over n^2} 
      \left( p_2 - {p_1\over 2} + \widetilde{p}_2 - 
                   {\widetilde{p}_1\over 2} \right) + {\cal O}(n^{-4}) 
      \right] \\
R_3(0,n,\rho) &=& {\theta_2 \over \theta_3} \left[ 1 + {1\over n^2}
      (\widetilde{p}_1 - \widetilde{p}_2 - p_3) + {\cal O}(n^{-4}) 
      \right] \\
R_4(0,n,\rho) &=& -\sinh (\rho\tau) \theta_2\theta_4  
                 \left[ 1 - {1\over n^2}  
      \left( {p_1\over 2} + \widetilde{p}_2 - {\widetilde{p}_1\over 2} 
             + p_3 \right) \right. \nonumber \\ 
     & & \qquad \left. \phantom{1\over n^2} + {\cal O}(n^{-4})
      \right] 
\label{ratios_R_final}
\end{subeqnarray} 

Plugging the formulae \reff{ratios_R_final} in the expression for the 
critical energy density \reff{energy_T=Tc} we get 
\be
-E_c(n,\rho) = \sqrt{2}  + {E_1(\rho) \over n}  + {E_3(\rho) \over n^3} 
   + {\cal O}\left({1\over n^5}\right)
\label{energy_final}
\end{equation}
where $E_1(\rho)$ \cite{Ferdinand_Fisher} and $E_3(\rho)$ are given by the 
expressions
\begin{eqnarray}
E_1(\rho)&=& { 2 \theta_2\theta_3\theta_4 \over \theta_2 + \theta_3 + \theta_4}
\label{def_E1} \\   
E_3(\rho)&=& -{ 2\theta_2\theta_3\theta_4 \over
              (\theta_2 + \theta_3 + \theta_4)^2} \left\{
      p_1 (\rho) \left( \theta_4 + {\theta_2 + \theta_3\over 2}\right) -
            p_2 (\rho) \theta_4  \right. \nonumber \\
  & & \qquad \qquad  + \left.
 \widetilde{p}_1(\rho) \left(            {\theta_2 - \theta_3\over 2}\right) +
 \widetilde{p}_2(\rho) \theta_3 +
            p_3 (\rho) \left(\theta_3 + \theta_4\right) \right\}
\label{def_E3}
\end{eqnarray}

The numerical values of the function $E_3(\rho)$ are given in 
Table~\ref{table_E3}

\begin{table}[tbh]
\centering
\begin{tabular}{rl}
\hline\hline 
\multicolumn{1}{c}{$\rho$}  & \multicolumn{1}{c}{$E_3(\rho)$} \\
\hline 
1        & $-$0.206683145336864 \\
2        & $-$0.184202899115749 \\
3        & $-$0.153247694215529 \\
4        & $-$0.102599506933675 \\
5        & $-$0.061201301359728 \\
6        & $-$0.034200082347112 \\
7        & $-$0.018369506074164 \\
8        & $-$0.009614465215356 \\
9        & $-$0.004941568941314 \\
10       & $-$0.002505707497764 \\
15       & $-$0.000074110828658 \\
20       & $-$0.000001946957522 \\
$\infty$ & \phantom{$-$}0 \\ 
\hline\hline 
\end{tabular}
\caption{\protect\label{table_E3}
Values of the coefficient $E_3(\rho)$ from \protect\reff{def_E3}
for several values of the torus aspect ratio $\rho$. 
}
\end{table}

\bigskip

\noindent
{\bf Remarks}. 1. After the completion of this work, Prof. Izmailian informed
us that the correct expression for the coefficient $E_3(\rho)$ had been 
published in the revised version of Ref.~\cite{Izmailian} (which can be
found in the Los Alamos preprint archive cond-mat). Their expression is
surprisingly simple
\be
E_3(\rho) = - {\pi^3 \rho \over 48} {\theta_2 \theta_3 \theta_4 \over 
             (\theta_2 + \theta_3 + \theta_4)^2} \, 
             \left[\theta_2^9 + \theta_3^9 + \theta_4^9 \right]  
\label{def_E3_Izmailian} 
\ee
Indeed, the numerical value of \reff{def_E3_Izmailian} coincides with our
result \reff{def_E3}. 
It would be interesting to find the analytic identities
proving the equivalence of \reff{def_E3} and \reff{def_E3_Izmailian}.

2. Let us check that \reff{def_E3} has the correct behavior under
$m \leftrightarrow n$ ($\rho \leftrightarrow 1/\rho$).
Indeed, the (trivial) fact that $E_c(n,m) = E_c(m,n)$ implies that  
we should have
\begin{subeqnarray}
\slabel{duality_E1}
E_1(\rho) &=& {E_1(1/\rho) \over \rho}  \\
E_3(\rho) &=& {E_3(1/\rho) \over \rho^3}
\slabel{duality_E3}
\end{subeqnarray}
The first equation \reff{duality_E1} can be easily proved by using Jacobi's 
imaginary transformation of the $\theta$-functions 
\reff{Jacobi_transformation}. Using these transformation one can easily show
that the second equation \reff{duality_E3} holds for \reff{def_E3_Izmailian}. 
We have also verified numerically that \reff{duality_E3} holds for our
result \reff{def_E3} to high accuracy using {\sc mathematica}. 

3. There is another simple way to test our results: We can first compute the 
{\em exact} value of the critical energy density $E_c(n,\rho)$ for several 
values of $n$ and a fixed value of $\rho$ by using 
\reff{def_energy}/\reff{def_z1}/\reff{table_notation}.
Then, we subtract the first two terms of the expansion 
\reff{energy_final} and fit the 
resulting function to the Ansatz $B_3 n^{-3} +  B_5 n^{-5} + B_7 n^{-7}$.
In Table~\ref{table_fits_energy} we show the numerical results for such fits
with $\rho=1, 2$, and 3. For $\rho=1$ we have used in the fits 1309 different
values between $n=16$ and $n=4096$. For $\rho=2,3$ we have used 10 different
values corresponding to $n=2^p$, $p=2,\ldots,11$ (that is why our 
estimates are more accurate for $\rho=1$ than for $\rho=2,3$).
We find an excellent agreement among 
the numerical estimates for $B_3$ and the exact values of $E_3$ quoted in 
Table~\ref{table_E3}. The value of $B_5$ for $\rho=1$ 
also agrees well with the value $\approx -0.7301823$ obtained by
Izmailian \cite{Izmailian_private} using analytic means.

\begin{table}[tbph]
\centering
\begin{tabular}{llll}
\hline\hline
$\rho$                & 
\multicolumn{1}{c}{1} & 
\multicolumn{1}{c}{2} & 
\multicolumn{1}{c}{3}\\
\hline
$B_3$  & $-0.2066831453369$  & $-0.1842028991157$ & $-0.1532476942155$ \\ 
$B_5$  & $-0.73018231235$    & $-0.416996817$     & $-0.316738073$     \\
$B_7$  & $-4.9362$           & $-3.405$           & $-2.693$           \\
\hline\hline
\end{tabular}
\caption{\protect\label{table_fits_energy}Fits of the function 
$-E_c(n,\rho) - \sqrt{2} - E_1(\rho)/n$  [cf.\ \protect\reff{def_E1}]
to the Ansatz 
$B_3 n^{-3} + B_5 n^{-5} + B_7 n^{-7}$ for several values of the 
torus aspect ratio $\rho$.
} 
\end{table}

4. In the limit $\rho\rightarrow\infty$ 
of an infinitely long torus (i.e.\ a cylinder) we have
\be
  E_1(\infty) = E_3(\infty) = 0
\label{energy_corrections_xi=infty}
\ee
as $\lim_{\rho\rightarrow\infty} \rho \theta_2 = 0$, and 
$\lim_{\rho\rightarrow\infty} \theta_3 = \lim_{\rho\rightarrow\infty} 
\theta_4 = 1$ [c.f.\ \reff{def_theta_functions}].

\bigskip

%
% SECTION 4
%
\section{Finite-size-scaling corrections to the specific heat}
\label{sec_specific_heat}

The goal of this section is to extend the asymptotic series of Ferdinand and 
Fisher \cite{Ferdinand_Fisher} for the specific heat through order $n^{-3}$. 
Let us start with the definition \reff{def_specific_heat} and see which terms
contribute to the critical value of $C_H$. The first term in 
\reff{def_specific_heat} is just a constant $(=-2)$; 
while the third term is quite similar to the one already obtained for 
the energy density \reff{energy_T=Tc_final}  
$[= - 4\rho R^{-2} R_4^2 \coth^2 (\tau\rho)]$. The most involved term is
the second one. Using the analysis of Ferdinand and Fisher, we can get the
final expression for the critical specific heat 
$C_{H,c}(n,\rho) = C_H(K_c,n,\rho)$:
\begin{eqnarray}
C_{H,c}(n,\rho) &=& 
    -2 + 4 Q_{1,-} - {4 R_3(0,n,\rho) \over R(0,n,\rho)} [ Q_{1,+} - Q_{1,-} ] 
        + 4 \rho {R_3(0,n,\rho) \over R(0,n,\rho)}
        \nonumber \\
    & & \qquad 
      - {4 \over R(0,n,\rho)} \sum\limits_{i=1}^3 R_i(0,n,\rho) Q_{1,i} 
         + {2 \sqrt{2} \over n} \lim\limits_{\tau\rightarrow0}  
        {R_4(\tau,n,\rho) \over R(0,n,\rho)} \coth \tau \rho \nonumber \\
    & & \qquad - 4 \rho \lim\limits_{\tau\rightarrow0} 
      \left( {R_4(\tau,n,\rho) \over R(0,n,\rho)} \coth \tau\rho
      \right)^2 
\label{specific_heat_T=Tc_final}
\end{eqnarray} 
where the $Q_{1,\pm}$ and $Q_{1,i}$ are those defined in 
Ref.~\cite{Ferdinand_Fisher} evaluated at $\tau=0$:
\begin{subeqnarray}
\slabel{def_Q11}
Q_{1,1}(n,\rho) &=& {1 \over n} \sum\limits_{r=0}^{n-1}
{1 - \tanh \left( {m \gamma_{2r+1} \over 2} \right) \over
\sin\left( {(r+1/2) \pi \over n}\right) \left[ 1 + \sin^2
    \left( {(r+1/2) \pi \over n}\right) \right]^{1/2} } \\
\slabel{def_Q12}
Q_{1,2}(n,\rho) &=& {1 \over n} \sum\limits_{r=0}^{n-1}
{1 - \coth \left( {m \gamma_{2r+1} \over 2} \right) \over
\sin\left( {(r+1/2) \pi \over n}\right) \left[ 1 + \sin^2
    \left( {(r+1/2) \pi \over n}\right) \right]^{1/2} } \\
\slabel{def_Q13}
Q_{1,3}(n,\rho) &=& {1 \over n} \sum\limits_{r=1}^{n-1}
{1 - \tanh \left( {m \gamma_{2r} \over 2} \right) \over
\sin\left( {r \pi \over n}\right) \left[ 1 + \sin^2
    \left( {r \pi \over n}\right) \right]^{1/2} } \\
\slabel{def_Q14}
Q_{1,4}(n,\rho) &=& {1 \over n} \sum\limits_{r=1}^{n-1} 
{1 - \coth \left( {m \gamma_{2r} \over 2} \right) \over 
\sin\left( {r \pi \over n}\right) \left[ 1 + \sin^2
    \left( {r \pi \over n}\right) \right]^{1/2} } \\  
\slabel{def_Q1-}
Q_{1,-}(n,\rho) &=& {1 \over n} \sum\limits_{r=0}^{n-1}
{1 \over
\sin\left( {(r+1/2) \pi \over n}\right) \left[ 1 + \sin^2
    \left( {(r+1/2) \pi \over n}\right) \right]^{1/2} } \\
Q_{1,+}(n,\rho) &=& {1 \over n} \sum\limits_{r=1}^{n-1}
{1 \over
\sin\left( {r \pi \over n}\right) \left[ 1 + \sin^2
    \left( {r \pi \over n}\right) \right]^{1/2} } 
\slabel{def_Q1+}
\label{def_Q}
\end{subeqnarray}

The terms with the factors $R_i/R$ ($i=3,4$) can be obtained easily using 
the results of Section~\ref{sec_energy}. Let us first consider 
the quantity $Q_{1,+}(n,\rho)$ \reff{def_Q1+}. The first step consists in 
expanding the factor $[1 + \sin^2(r\pi/n)]^{-1/2}$ in Eq.~\reff{def_Q1+}
in power series of $\sin(r \pi/n)$:
\begin{eqnarray}
\label{Q1+_1}
Q_{1,+} &=& {1\over n}\sum\limits_{r=1}^{n-1} {1 \over \sin\left(
            {r \pi \over n} \right) } + 
            {1\over n}\sum\limits_{r=1}^{n-1} \sum\limits_{k=1}^\infty
            \left(\begin{array}{c}
                    -1/2 \\
                     k 
                   \end{array}\right) \sin^{2k-1} 
             \left({r \pi \over n} \right) \\
        &\equiv& Q_{1,+}^{(1)} + Q_{1,+}^{(2)}
\label{Q1+_2}
\end{eqnarray} 
The computation of $Q_{1,+}^{(2)}$ is done by applying the Euler-MacLaurin 
formula \reff{Euler_MacLaurin_formula} to the function $f(p)=\sin^{2k-1}(p)$
with $L=2n$ and $\alpha=1/2$. The result is 
\be
Q_{1,+}^{(2)}(n,\rho) = -{\log 2 \over \pi} + {\pi \over 12} {1 \over n^2} 
  + {\cal O}(n^{-4}) 
\label{Q1+2_final}
\ee
The computation of the divergent part $Q_{1,+}^{(1)}$ is a little 
more involved. The idea
if to apply the Euler-MacLaurin formula \reff{Euler_MacLaurin_formula} to
the function $f(p) = \sin^{-1}(p) -1/p + 1/(p-\pi)$ with $L=2n$ and 
$\alpha=1/2$:
\begin{eqnarray}
  {1\over n} \sum\limits_{r=0}^{n-1} \left[ 
     \sin^{-1}\left( {r \pi \over n}\right) - {n \over r \pi} + 
                   {n \over \pi(r-n) } \right]  &=& 
{2\over \pi} \log {2\over \pi} + {\pi \over 6n^2}
                 \left( {1\over \pi^2} - {1 \over 6} \right)
              + {\cal O}(n^{-4})  \nonumber \\
 &=&  Q_{1,+}^{(1)} - {2\over \pi} \sum\limits_{r=1}^{n-1}
               {1\over r} - {1 \over \pi n} 
\label{serie_cv}
\end{eqnarray}
Using the well-known asymptotic expansion
%%% \cite{Caracciolo_98} [See also \reff{sum_one_over_n}]
\cite{GR} [See also \reff{sum_one_over_n}]
\be
 \sum\limits_{r=1}^{L} {1\over r} = \log L + \gamma_E 
  + {1 \over 2L} - {1\over 12 L^2} + {\cal O}(L^{-4}) 
\ee
(where  $\gamma_E \approx 0.5772156649$ is the Euler constant) we finally get
\be
Q_{1,+}^{(1)}(n,\rho) = {2 \over \pi} \left[ \log n + \gamma_E + 
\log {2 \over \pi} - {\pi^2 \over 72 n^2} + {\cal O}(n^{-4}) \right] 
\label{Q1+1_final}
\ee
Putting together \reff{Q1+2_final}/\reff{Q1+1_final} we arrive at the
final result
\be
Q_{1,+}(n,\rho) = {2 \over \pi} \left[ \log n + \gamma_E + 
                 \log {2^{1/2} \over \pi} + {\pi^2 \over 36 n^2} 
                 + {\cal O}(n^{-4}) \right] 
\label{Q1+_final}
\ee
Using similar methods we obtain\footnote{The leading term of this equation 
was obtained by Onsager \protect\cite{Onsager}.
}
\be
Q_{1,-}(n,\rho) = {2 \over \pi} \left[ \log n + \gamma_E + 
                 \log {2^{5/2} \over \pi} - {\pi^2 \over 72 n^2} 
                 + {\cal O}(n^{-4}) \right] 
\label{Q1-_final}
\ee

The last part consists in evaluation the $Q_{1,i}$ (with $i=1,2,3$) 
in \reff{def_Q}. For brevity we will do explicitly the simplest case 
$Q_{1,4}(n,\rho)$ \reff{def_Q14}. The first step is to expand 
the term $1 - \coth (m\gamma_{2r}/2)$ as a power series in 
$\exp(-m\gamma_{2r})$  
\be
Q_{1,4}(n,\rho) = {2\over n} \sum\limits_{r=1}^{\lfloor{n\over 2}\rfloor} 
                   \sum\limits_{p=1}^\infty {e^{-m p \gamma_{2r} } \over 
       \sin\left( {r \pi \over n}\right) \sqrt{ 1 + 
       \sin^2\left( {r \pi \over n}\right) } } 
\label{Q14_1}
\ee
Then we split the sum over $r$ as in \reff{sum_splitting_trick}. The term
including the sum $\sum_{r=s(n)}^{[n/2]}$ gives a total contribution of
order $n^{-(M-1)}$ with the choice \reff{def_s} for $s(n)$. The second step
is to plug into \reff{Q14_1} the expansion \reff{series_m_gamma_2r} for
$m \gamma_{2r}$
\begin{subeqnarray}
Q_{1,4} &=& - {4 \over n} \sum\limits_{p=1}^\infty \sum\limits_{r=1}^{s-1}
              {e^{-2 \pi r p \rho} \over
                 \sin\left( {r \pi \over n}\right) \sqrt{ 1 +
                 \sin^2\left( {r \pi \over n}\right) } } \nonumber \\ 
        & & \qquad 
            - {8 \pi^3 \rho \over 3 n^3} 
              \sum\limits_{p=1}^\infty p \sum\limits_{r=1}^{s-1} r^3 
             {e^{-2 \pi r p \rho} \over
                 \sin\left( {r \pi \over n}\right) \sqrt{ 1 +
                 \sin^2\left( {r \pi \over n}\right) } } + 
             {\cal O}(n^{-4}) \\
        &\equiv& Q_{1,4}^{(a)} + Q_{1,4}^{(b)}   
\end{subeqnarray} 

The computation of $Q_{1,4}^{(b)}$ is quite easy: we expand the factors
$\sin(r \pi/n)$ in powers of $r \pi/n$:
\be
Q_{1,4}^{(b)} = - {8 \pi^2 \rho \over 3 n^2}
              \sum\limits_{p=1}^\infty p \sum\limits_{r=1}^{s(n)-1} r^2
             e^{-2 \pi r p \rho} +  
             {\cal O}(n^{-4}) 
\label{Q14b_1}
\ee  
Then we can extend the sum $\sum_{r=1}^{s(n)-1}$ to $\sum_{r=1}^\infty$ 
at an error of order $n^{-(M+2)} \log^2 n$. Using the fact that
\be
\sum\limits_{p=1}^\infty p e^{-2\pi r p \rho} = { e^{-2\pi r\rho} \over  
              \left( 1 - e^{-2\pi r\rho} \right)^2 } = 
                 {1 \over 4 \sinh^2(\pi r \rho)} 
\label{sum_p}
\ee
we have that 
\be
Q_{1,4}^{(b)} = - {2 \pi^2 \rho \over 3 n^2} 
                \sum\limits_{p=1}^\infty {r^2 \over \sinh^2(\pi r \rho)} 
\label{Q14b_final}
\ee

The computation of $Q_{1,4}^{(a)}$ follows the same steps:
\be
Q_{1,4}^{(a)} = - {4 \over \pi}
              \sum\limits_{p=1}^\infty p \sum\limits_{r=1}^\infty {1 \over r}
              e^{-2 \pi r p \rho} +  
                {4 \pi \over 3 n^2 } 
              \sum\limits_{p=1}^\infty p \sum\limits_{r=1}^\infty r 
             e^{-2 \pi r p \rho} 
             + {\cal O}(n^{-4})
\label{Q14a_1}
\ee
In this case the error introduced by extending the sum $\sum_{r=1}^{s(n)-1}$ 
to $\sum_{r=1}^\infty$ is of order $n^{-M}$. Using \reff{sum_p} we 
find that
\be
Q_{1,4}^{(a)} =  {2 \over \pi}
              \sum\limits_{r=1}^\infty {1 \over r} [ 1 - \coth( \pi r \rho) ]
               - {2 \pi \over 3 n^2 }
               \sum\limits_{r=1}^\infty r [ 1 - \coth (\pi r \rho) ] 
             + {\cal O}(n^{-4})
\label{Q14a_final}
\ee
Thus, we write the final result as
\begin{subeqnarray}
\slabel{Q14_final}
Q_{1,4}(n,\rho) &=& Q_{1,4}^{(0)}(\rho) + {Q_{1,4}^{(2)}(\rho) \over n^2} 
                 + {\cal O}(n^{-4}) \\[3mm]
\slabel{Q140_final}
Q_{1,4}^{(0)}(\rho) &=& {2 \over \pi} 
\sum\limits_{r=1}^\infty {1 \over r} [ 1 - \coth( \pi r \rho) ] \\
Q_{1,4}^{(2)}(\rho) &=& -{2\pi\over 3} \left\{ 
         \sum\limits_{r=1}^\infty r [ 1 - \coth (\pi r \rho) ]  + \pi \rho 
         \sum\limits_{r=1}^\infty r^2 \left[ i\, {\rm csch}(\pi r \rho) 
         \right]^2 \right\} 
\slabel{Q142_final}
\label{Q14_final_all}
\end{subeqnarray}
The other three quantities $Q_{1,i}$ (\ref{def_Q11}-c) 
can be computed in a similar way.
The result can be written as \reff{Q14_final_all} using the translations
given by \reff{table_notation} [In this case, the third column of 
\reff{table_notation} does not play any role].
For $i=1,2$ we should make two slight 
modifications: (a) The factor $r$ in \reff{Q14_final_all} should
be replaced by $r+1/2$, and (b) the sums over $r$ in $Q_{1,1}$ and $Q_{1,2}$
start at $r=0$ rather than at $r=1$ (as in $Q_{1,3}$ and $Q_{1,4}$). 

Putting all the pieces together we arrive at the final result
\be
C_{H,c}(n,\rho) = 
                 {8 \over \pi} \log n + C_0(\rho) + {C_1(\rho) \over n} + 
                 {C_2(\rho) \over n^2} + {C_3(\rho) \over n^3} +
                 {\cal O}\left( {1\over n^4}\right)
\label{specific_heat_final}
\ee
where the coefficients $C_i(\rho)$ are given by 
\begin{eqnarray}
\label{def_C0}
C_0(\rho) &=& {8 \over \pi} \left( 
      \log {2^{5/2} \over \pi} + \gamma_E - {\pi \over 4} \right) \nonumber \\ 
         & & \qquad 
      - {4 \over \theta_2 + \theta_3 + \theta_4}
        \left[ {4 \over \pi} \sum\limits_{\nu=2}^4 \theta_\nu \log \theta_\nu
              + \rho { \theta_2^2 \theta_3^2 \theta_4^2  \over
                 \theta_2 + \theta_3 + \theta_4} \right] \\
\label{def_C1}
C_1(\rho) &=&  - 2 \sqrt{2} {\theta_2\theta_3\theta_4 \over
          \theta_2 + \theta_3 + \theta_4} = - \sqrt{2} E_1(\rho) \\ 
\label{def_C2}
C_2(\rho) &=&   - 4 \rho { \theta_2\theta_3\theta_4 \over
                       \theta_2 + \theta_3 + \theta_4} E_3(\rho) - 
               {\pi \over 9} 
             - 4 A_3(\rho) {\theta_2 \over \theta_2 + \theta_3 + \theta_4} 
               \left(\rho - {\log 16 \over \pi} \right) \nonumber \\
    & & \qquad 
         + {\pi \over 3} {\theta_2 \over \theta_2 + \theta_3 + \theta_4} - 
         {16 \over \pi} \left[ A_2(\rho) \theta_4 + A_3(\rho) \theta_2 \right] 
         {\sum\limits_{\nu=2}^4 \theta_\nu \log \theta_\nu \over 
          (\theta_2 + \theta_3 + \theta_4)^2 } \nonumber \\
   & & \qquad 
       - {4 \over \theta_2 + \theta_3 + \theta_4} G(\rho) \\
C_3(\rho) &=& - \sqrt{2} E_3(\rho) 
\label{def_C3}
\end{eqnarray}
where 
\begin{subeqnarray} 
G(\rho) &=& Q_{1,1}^{(2)} \theta_3 + Q_{1,2}^{(2)} \theta_4 + 
         Q_{1,3}^{(2)} \theta_2 - Q_{1,2}^{(0)} A_2(\rho) \theta_4 - 
         Q_{1,3}^{(0)} A_3(\rho) \theta_2 \\ 
A_2(\rho) &=&  p_2(\rho) + \widetilde{p}_2(\rho) - {1\over 2}\left[ 
             p_1(\rho) + \widetilde{p}_1(\rho) \right] \\
A_3(\rho) &=& p_3(\rho) + \widetilde{p}_2(\rho) - p_1(\rho) 
\end{subeqnarray} 
The expressions for $C_0$ and $C_1$ were first obtained by Ferdinand and Fisher
\cite{Ferdinand_Fisher}. The results for $C_2$ and $C_3$ are new. In 
Table~\ref{table_C2} we show the values of the coefficient $C_2(\rho)$ for 
several values of the aspect ratio $\rho$.

\begin{table}[tbh]
\centering
\begin{tabular}{rl}
\hline\hline 
\multicolumn{1}{c}{$\rho$}  & \multicolumn{1}{c}{$C_2(\rho)$} \\
\hline 
1  & \phantom{$-$}0.097119896855337 \\
2            & $-$0.326865280829340 \\  
3            & $-$0.748187561687100 \\
4            & $-$0.877385104391125 \\
5            & $-$0.809168407448959 \\
6            & $-$0.682469414146146 \\
7            & $-$0.567445479079586 \\
8            & $-$0.483401539198706 \\
9            & $-$0.428249598449714 \\
10           & $-$0.394311952593824 \\
15           & $-$0.351150319692501 \\
20           & $-$0.349140134316672 \\
$\infty$     & $-$0.349065850398866 \\ 
\hline\hline 
\end{tabular}
\caption{\protect\label{table_C2}
Values of the coefficient $C_2(\rho)$ from \protect\reff{def_C2}
for several values of the torus aspect ratio $\rho$. 
}
\end{table}

\bigskip

\noindent
{\bf Remarks} 1. As shown by Ferdinand and Fisher using the Jacobi's 
transformations \reff{Jacobi_transformation}, the coefficient $C_0(\rho)$ 
satisfies the identity
\be
C_0(\rho) = C_0(\rho^{-1}) + {8\over \pi} \log \rho
\label{duality_C0}
\ee
Indeed, from relations \reff{def_C1}/\reff{def_C3} we conclude that
these two coefficients have the right behavior under the transformation
$\rho \rightarrow 1/\rho$ [c.f.,\reff{duality_E1}/\reff{duality_E3}]. 
Finally, we have tested numerically that 
\begin{equation} 
C_2(\rho) = {C_2(1/\rho) \over \rho^2} 
\label{duality_C2}
\end{equation}
is satisfied. This is a non-trivial test of the correctness of our result. 

2. We have also performed the following test: we have defined the function  
equal to the {\em exact} value of the specific heat $C_{H,c}(n,\rho)$ 
minus the first three terms of the expansion \reff{specific_heat_final}. 
Then, we have fitted the result to several Ans\"atze. In particular, 
we show in Table~\ref{table_fits_specific_heat} the results for the Ansatz
$D_2/n^2 + D_3/n^3 + D_4/n^4 + D_5/n^5$. We have used the same data as 
for the energy fits in Section~\ref{sec_energy}. 
The agreement between the values $D_2,D_3$ and the exact values 
\reff{def_C2}/\reff{def_C3}/\reff{def_E3} is very good.  
It is also interesting to note 
that $D_5(\rho) \approx - \sqrt{2} B_5(\rho)$, where $B_5(\rho)$ is the 
coefficient obtained in a similar fit to the energy 
(see Table~\ref{table_fits_energy}).

\begin{table}[tbph]
\centering
\begin{tabular}{llll}
\hline\hline
$\rho$                & 
\multicolumn{1}{c}{1} & 
\multicolumn{1}{c}{2} & 
\multicolumn{1}{c}{3}\\
\hline
$D_2$  & $0.097119896855$  & $-0.3268652808$ & $-0.7481875617$ \\
$D_3$  & $0.2922941107$    & $\phantom{-}0.2605022$ & $\phantom{-}0.2167250$ \\
$D_4$  & $0.014792$        & $-1.0635$       & $-1.8668$ \\
$D_5$  & $1.0326$          & $\phantom{-}0.59$      & $\phantom{-}0.45$ \\
\hline\hline
\end{tabular}
\caption{\protect\label{table_fits_specific_heat}Fits of the function
$C_{H,c}(n,\rho) - (8/\pi) \log n - C_0(\rho) - C_1(\rho)/n$ 
[cf.\ 
\protect\reff{specific_heat_final}/\protect\reff{def_C0}/\protect\reff{def_C1}]
to the Ansatz $D_2 n^{-2} + D_3 n^{-3} + D_4 n^{-4} + D_5 n^{-5}$ for several 
values of the torus aspect ratio $\rho$.
}
\end{table}

3. By inspection from Table~\ref{table_C2}, the function $C_2(\rho)$ should
have a zero at a non-trivial value of the aspect ratio $\rho_{\rm min}$ 
between 1 and 2. We have evaluated numerically that value
\be
\rho_{\rm min} \approx 1.33544086
\label{xi_C2=0}
\ee
By \reff{duality_C2}, there is another zero of $C_2(\rho)$  
at $\rho_{\rm min}^{-1} \approx 0.74881639$.

4. In the limit $\rho\rightarrow\infty$ the coefficients $C_i$ tend to the 
following limits:
\begin{subeqnarray}
C_0(\infty) &=& {8 \over \pi} \left(
                \log {2^{5/2} \over \pi} + \gamma_E - {\pi \over 4} \right)\\
C_1(\infty) &=& 0 \\
C_2(\infty) &=& -{\pi \over 9} \\
C_3(\infty) &=& 0
\label{corrections_spacific_heat_xi=infty}
\end{subeqnarray} 
Thus, only the coefficients associated to even powers of $n^{-1}$ survive in 
this limit. 

\bigskip

%
% SECTION 5
%
\section{Further remarks and conclusions} \label{sec_conclusions}

The computation of the finite-size corrections to the
energy and specific heat shows
that, through order $n^{-3}$: 

\begin{itemize}

\item[(a)] All the corrections are integer powers of the quantity $n^{-1}$. In 
   particular there are no multiplicative or additive logarithmic terms 
   (except for the leading term in the specific heat). 

\item[(b)] In the energy density we only find {\em odd} powers of $n^{-1}$.
   In particular, the corrections of order $n^{-2}$ and $n^{-4}$ are absent  
   in \reff{energy_final}. 

\item[(c)] In the specific heat we find both even and odd powers of 
    $n^{-1}$.  But the coefficients of the corrections corresponding
    to odd powers of $n^{-1}$ in \reff{specific_heat_final} are proportional 
    to the corresponding coefficients in the energy expansion 
    \reff{energy_final}:
    we found that $C_i(\rho) = - \sqrt{2} E_i(\rho)$ 
    for $i=1,3$ [cf.\ \reff{def_C1}/\reff{def_C3}],
    and the numerical test performed at the end of 
    Sections~\ref{sec_energy} and \ref{sec_specific_heat} shows that the 
    same ratio holds for the next coefficients $C_5/E_5 \approx -\sqrt{2}$.  

\end{itemize}
The natural question is whether these observations are general features
that hold to all orders in $n^{-1}$. In this section, we 
will try to answer those questions.

Let us first analyze what happens to the energy. We first note that in the  
expansion \reff{series_m_gamma_2r} only even powers of $n^{-1}$ occur.
This expansion can be done to any finite order we want. The errors coming from
neglecting the second sum in \reff{sum_splitting_trick} and from extending
the sums $\sum_{r=1}^{s(n)-1}$ to $\sum_{r=1}^\infty$ are at most of order 
${\cal O}(n^{-(M-1)})$, and they can be made as small as we want by making $M$ 
in the definition of $s(n)$ \reff{def_s} as large as we need. 
This means that in the series expansions of the quantities $\log P_i$ 
($i=1,\ldots,4$) \reff{P's_final} only even powers of $n^{-1}$ appear.
Secondly, we need to check that only even powers of $n^{-1}$ occur in 
the expansion of $\log P_0$ \reff{logP0_1_1}. The argument is simple: all 
derivatives $f^{(k)}$ of the function $f(p) = \log(\sin p + \sqrt{1+\sin^2 p})$
are integrable over the interval $[0,\pi]$. This implies that the 
Euler-MacLaurin formula \reff{Euler_MacLaurin_formula} can be applied to any 
arbitrary order. As $f(0) = f(\pi)$, the correction of order $n^{-1}$ 
vanishes and only corrections with even powers of $n^{-1}$ can occur. 
The same conclusion applies to the expansion \reff{logP0_1_1} and to the
ratios $R_i$ \reff{ratios_R_final}. Thus, from formula \reff{energy_T=Tc_final}
we immediately conclude that only {\em odd} powers of $n^{-1}$ appear  
in the finite-size corrections to the critical energy density. This 
result generalizes point (b) above. In particular, no logarithmic corrections
occur in this expansion at any order.

\bigskip

\noindent
{\bf Remark}. The authors of Ref.~\protect\cite{Izmailian} made an argument 
to explain the absence of the $L^{-2}$ correction in the energy. 
They started with the Fortuin--Kasteleyn representation of the $q$-state 
Potts model on a graph $G$ ($q=2$ corresponds to the Ising model) 
\cite{FK1,FK2}: 
\be  
Z_G(K) = \sum\limits_{G'\subseteq G} v^{N_b(G')} q^{N_c(G')} \; ,  
\ee
where $N_b$ and $N_c$ are respectively the number of bonds and connected 
clusters of the spanning subgraph $G'\subseteq G$, and $v=e^{K} -1$. 
Then, they introduced the standard finite-size-scaling Ansatz for the free 
energy assuming there are no irrelevant operators \cite{Privman}: 
\be
f_G(K) = f_{\rm reg}(K) + {1 \over L_x L_y} W(L_x^{1/\nu}(K-K_c)) \,. 
\label{FSS_Ansatz}
\ee
Here $G$ is a square lattice of size $L_x \times L_y$, $f_{\rm reg}(K)$ 
is the regular part of the free energy, $\nu$ is the 
usual critical exponent, and $W(x)$ is an analytic function at $x=0$. 
They expanded $W(x)$ around $x=0$ and computed the mean values of  
$N_c$ and $N_b$ at criticality ($K=K_c$). They found that 
\begin{subeqnarray}
\< N_c \> &=& n_c L_x L_y + A L_x^{1/\nu} + B \\
\< N_b \> &=& n_b L_x L_y + C L_x^{1/\nu} 
\end{subeqnarray} 
with no constant correction to $\< N_b \>$. The internal energy 
\reff{def_energy} is related linearly with $N_b$; thus, in the 
Ising model at criticality, we have $E_c = A_0 + A_1(\rho) L_x^{-1}$ with 
no higher-order corrections in $L_x^{-1}$. They concluded 
that the $L_x^{-2}$ correction to the critical energy should vanish.  
Indeed, this argument also implies  
the stronger result that there are no corrections of any kind beyond order 
$L_x^{-1}$ --- a conclusion that is unfortunately false! 
So it is not clear that the Ansatz \reff{FSS_Ansatz} (even if we include 
irrelevant operators) is enough to reproduce the right finite-size expansion 
of the internal energy of the critical two-dimensional Ising model  
\reff{energy_final}. 
A detailed discussion of this point will be published elsewhere 
\protect\cite{Salas_Sokal_in_prep} (See also
\protect\cite[Section~3]{Salas_Sokal_Ising} for preliminary results).

\medskip

The analysis of the specific heat is a little more involved. Using the 
same procedure as for the energy, we conclude that the fourth ($\sim R_3/R$) 
and the last ($\sim R_4^2/R^2$) terms of \reff{specific_heat_T=Tc_final} 
provide corrections with {\em even} powers of $n^{-1}$. Furthermore, the  
sixth term of \reff{specific_heat_T=Tc_final} [$\sim R_4/(R n)$] will
give only {\em odd} powers of $n^{-1}$.

Let us see nor what happens to the
terms $Q_{1,i}$ (\ref{def_Q11}-d). The argument for $Q_{1,4}$ \reff{def_Q14} 
is quite simple: the expansion of this quantity as a power series of 
$r\pi/n$ will only contain even powers of $n^{-1}$. Indeed, the errors 
coming from dropping the sum $\sum_{r=s(n)}^{\lfloor n/2 \rfloor}$ in the 
beginning 
of the computation, and for extending the sum $\sum_{r=1}^{s(n)-1}$ to 
$\sum_{r=1}^\infty$ at the end of the computation are both at most of 
order ${\cal O}(x^{-(M-1)})$ with the choice \reff{def_s} for $s(n)$. 
Thus, the fifth term in \reff{specific_heat_T=Tc_final} 
[$\sim \sum R_i Q_{1,i}/R$]
will have only corrections with {\em even} powers of $n^{-1}$.

Finally, let us analyze the behavior of $Q_{1,+}$ \reff{def_Q1+}. In the
evaluation of $Q_{1,+}^{(2)}$ we had to apply the Euler-MacLaurin 
formula \reff{Euler_MacLaurin_formula} to the function $f(p) = \sin^{2k-1} p$
with $k \geq 1$. The derivatives of this function are always integrable 
over the interval $[0,\pi]$, thus the expansion \reff{Q1+2_final} only 
contains {\em even} powers of $n^{-1}$. In the evaluation of 
$Q_{1,+}^{(0)}$ we apply the Euler-MacLaurin formula 
\reff{Euler_MacLaurin_formula} to the function 
$f(p) = 1/\sin p - 1/p + 1/(p-\pi)$. This function and all its derivatives
are integrable over the interval $[0,\pi]$. Thus, Eq.~\reff{serie_cv} can
be generalized to
\be
Q_{1,+}^{(0)} = {2\over \pi}\left\{ \sum\limits_{p=1}^{n-1} {1\over p} 
             + {1\over 2n} + \log {2 \over \pi} + 
             {\pi \over 2} \sum\limits_{k=1}^\infty {B_{2k} \over 2k} 
             \left({\pi \over n}\right)^{2k} 
             \left[ f^{(2k-1)}(\pi) - f^{(2k-1)}(0) \right] \right\} 
\label{serie_cv_1}
\ee
This expansion contains in principle both even and odd powers of $n^{-1}$. 
We now plug in the well-known result \cite{GR}
%%% \cite{Caracciolo_98}
%
\be
\sum\limits_{p=1}^{L} {1 \over p} = \log L + \gamma_E + {1 \over 2L } 
        - \sum\limits_{k=1}^\infty {B_{2k} \over 2k} L^{-2k} 
\label{sum_one_over_n}
\ee
with $L=n-1$, and expand the resulting factors $\log(1 -1/n)$, $(1-1/n)^{-1}$, 
and $(1-1/n)^{-2k}$ in powers of $n^{-1}$. If we express the result as 
$\sum_k \alpha_k n^{-k}$, we immediately see that $\alpha_1=0$. The
expression for the coefficients of the odd powers of $n^{-1}$ is given by 
\be
\alpha_{2k+1} = {1\over 2} - {1\over 2k+1} - \sum\limits_{m=1}^{k} 
           \left( \begin{array}{c}
                  2k \\
                 2m-1 
                 \end{array} \right) {B_{2m} \over 2m}\, , \qquad k\geq 1 
\label{formula_alpha}
\ee
To prove $\alpha_{2k+1}=0$, we can apply the general Euler-MacLaurin 
formula \reff{Euler_MacLaurin_general} to the function $f(x)=x^{2k}$ with 
$n=0$ and $m=1$:
\begin{eqnarray}
0 &=& \int_0^1 x^{2k} \, dx - {1\over 2} 
                  + \sum\limits_{m=1}^{k} {B_{2m} \over (2m)!} 
                  { (2k)! \over (2k-2m+1)! }  = -\alpha_{2k+1} 
\end{eqnarray}
This implies that in the finite-size-scaling 
expansion of $Q_{1,+}$ only {\em even} powers of $n^{-1}$ occur. The same
holds for $Q_{1,-}$ \reff{def_Q1-}.

In summary, we have seen that in the finite-size-scaling expansion of the 
specific heat at criticality only integer powers of $n^{-1}$ appear
[except of course for the leading term $(8/\pi)\log n$]. Furthermore, 
all contributions to the specific heat \reff{specific_heat_T=Tc_final} give 
{\em even} powers of $n^{-1}$, except for one: namely, the sixth term in 
\reff{specific_heat_T=Tc_final}, which has the form
\be
{2 \sqrt{2} \over n} \lim\limits_{\tau\rightarrow0} 
{R_4(\tau,n,\rho) \over R(0,n,\rho)} \, \coth \tau \rho 
\label{one_over_n_cv}
\ee   
If we compare it to the expression for the energy \reff{energy_T=Tc_final} we
conclude that the coefficients associated to {\em odd} powers of $n^{-1}$
in the energy and specific-heat expansions are proportional. In particular 
[c.f., \reff{expansion_energy}/\reff{expansion_specific_heat}], and recalling  
that $E_{2k}=0$, we obtain 
\be
{E_k(\rho) \over C_k(\rho) } = \left\{ \begin{array}{ll} 
                            - 1/\sqrt{2} &  \qquad \hbox{\rm for $k$ odd}\\
                              0          &  \qquad \hbox{\rm for $k$ even}
                                       \end{array}
                               \right.   
\label{proportionality}
\ee
This generalizes the results \reff{def_C1}/\reff{def_C3}. Note in 
particular that the ratio $E_k(\rho)/C_k(\rho)$ is independent of the
aspect ratio $\rho$. 

Let us conclude with a brief discussion about the generality of the above 
results. It is clear that the leading-term coefficients $E_0$ and $C_{00}$ 
in \reff{expansion_energy}/\reff{expansion_specific_heat} do {\em not} depend
on the boundary conditions, as they are bulk quantities. However, the   
finite-size-scaling coefficients $E_k(\rho)$ and $C_k(\rho)$ with 
$k\geq 1$ are expected to depend in general on the boundary conditions 
of the system. 
In particular, Lu and Wu \cite{Lu} have obtained the finite-size expansion 
of the free energy for a critical Ising model with two different boundary 
conditions (namely, a M\"obius strip and a Klein bottle). They found an 
expansion of the form 
\be
f_c(n,\rho) = f_0 + {f_1(\rho) \over n } + {f_2(\rho) \over n^2 } + \cdots  
\label{expansion_free_energy}
\ee
where the coefficients $f_k(\rho)$ with $k=1,2$ {\em do} depend 
explicitly on the boundary conditions.

On the other hand, the fact that the ratio $E_k(\rho)/C_k(\rho)$ is a 
$\rho$-independent number suggests that it might be universal, i.e., 
independent of the details of the Hamiltonian. It would be very interesting
to investigate this possibility. 
A first step in this direction has been achieved by Izmailian and Hu 
\cite{Izmailian3} who computed the finite-size expansion of the free energy 
per spin $f_N$ and the inverse correlation length $\xi^{-1}_N$ for a critical 
Ising model on  several $N\times\infty$ lattices with periodic boundary 
conditions (namely, square, hexagonal and triangular). On each lattice they 
found expansions of the form 
\begin{subeqnarray}
f_N - f_0  &=& \sum\limits_{k=1}^\infty {f_k \over N^{2k} } \\
\xi^{-1}_N &=& \sum\limits_{k=1}^\infty {b_k \over N^{2k-1} }  
\end{subeqnarray} 
where the coefficients $b_k$ and $f_k$ {\em do} depend on the lattice. 
However, their ratio $b_k/f_k$ is the same for all three lattices and equal to  
$b_k/f_k = (2^{2k}-1)/(2^{2k-1}-1)$. 
They also computed the corresponding expansions for a quantum spin chain
belonging to the two-dimensional Ising model universality class. They found 
that the ratio $b_k/f_k$ takes the same value as in the Ising case. This 
result supports the conjecture that this ratio is a universal quantity.

\appendix

%
% APPENDIX A 
%
\section{Theta functions} \label{sec_theta}

In this appendix we gather all the definitions and properties of the 
Jacobi's $\theta$-functions needed in this paper. We follow the notation
of Ref.~\cite{Ferdinand_Fisher}, which was adapted from 
Whittaker and Watson \cite{Whittaker}. 
We define the Jacobi $\theta$-functions $\theta_i$ at $z=0$ in the following 
way:\footnote{In this particular case, $\theta_1(0,e^{-\pi\rho})=0$.
}
\begin{subeqnarray}
\theta_2 &\equiv& \theta_2(0,e^{-\pi\rho}) = 2\theta_0 e^{-\pi\rho/4} 
                  \prod\limits_{r=1}^\infty 
                  \left( 1 + e^{-2r \pi \rho}\right)^2 \\
\theta_3 &\equiv& \theta_3(0,e^{-\pi\rho}) = \theta_0 
                  \prod\limits_{r=1}^\infty 
                  \left( 1 + e^{-(2r-1)\pi \rho}\right)^2 \\
\theta_4 &\equiv& \theta_4(0,e^{-\pi\rho}) = \theta_0 
                  \prod\limits_{r=1}^\infty 
                  \left( 1 - e^{-(2r-1)\pi \rho}\right)^2 
\label{def_theta_functions}
\end{subeqnarray}
The function $\theta_0(\rho)$ is defined as
\be
\theta_0 = \theta_0(\rho) =
           \prod\limits_{r=1}^\infty \left( 1 - e^{-2\pi r\rho} \right)
\label{def_theta0}
\ee
and it satisfies the following identity
\be
\theta_0 = e^{\pi \rho/12} \left[ {1\over 2} \theta_2 \theta_3 \theta_4
                          \right]^{1/3}
\label{property_theta0}
\ee
These $\theta$-functions \reff{def_theta_functions} satisfy the Jacobi's 
imaginary transformation
\begin{subeqnarray}
\theta_2\left(0,e^{-\pi/\rho}\right) &=& \rho^{1/2}
                                     \theta_4\left(0,e^{-\pi\rho}\right)\\
\theta_3\left(0,e^{-\pi/\rho}\right) &=& \rho^{1/2}
                                     \theta_3\left(0,e^{-\pi\rho}\right)\\
\theta_4\left(0,e^{-\pi/\rho}\right) &=& \rho^{1/2}
                                     \theta_2\left(0,e^{-\pi\rho}\right)
\label{Jacobi_transformation}
\end{subeqnarray}
%

%
% APPENDIX B 
%
\section{The Euler-MacLaurin formula} \label{sec_Euler}

The Euler-MacLaurin formula (see e.g.\ \cite[Appendix B]{Caracciolo_98})
is the main tool we need to compute asymptotic series. 
The general form of this formula is given by 
\begin{eqnarray}
\sum\limits_{k=n}^{m-1} f(k) &=& \int_n^m dx f(x) - {1 \over 2}[f(m) - f(n)] 
        \nonumber \\
  & & \qquad
+ \sum\limits_{p=1}^N {B_{2p}\over (2p)!} [f^{(2p-1)}(m) - f^{(2p-1)}(n)]
  \nonumber \\
  & & \qquad + {1\over (2N+1)!} 
   \int_n^m dx f^{(2N+1)}(x) B_{2N+1}(x - \lfloor x\rfloor)
\label{Euler_MacLaurin_general}
\end{eqnarray}
where $B_n$ are the Bernoulli numbers and $B_n(x)$ are the Bernoulli 
polynomials defined by 
\be
B_n(x) = \sum\limits_{k=0}^n \left(\begin{array}{l}
                                     n \\
                                     k
                                   \end{array} \right) B_k x^{n-k}
\ee

In this paper we are mainly interested in sums of the form
\be
{1 \over L} \sum\limits_{n=0}^{\alpha L -1 } f(p) 
\label{goal_sum}
\ee
where $p = 2\pi n/L$. The asymptotic expansion of the sum \reff{goal_sum}
in the limit $L\rightarrow\infty$ with $\alpha$ fixed can be obtained from 
\reff{Euler_MacLaurin_general}: 
\begin{eqnarray}
{1 \over L} \sum\limits_{n=0}^{\alpha L-1} f(p) &=& \int_0^{2\pi\alpha}
{dp \over 2\pi} f(p) - {1 \over 2L}[f(2 \pi \alpha) - f(0)] \nonumber \\
  & & \qquad
+ {1\over 2\pi} \sum\limits_{k=1}^N {B_{2k}\over (2k)!} \left(
  {2 \pi \over L} \right)^{2k} [f^{(2k-1)}(2\pi\alpha) - f^{(2k-1)}(0)]
  \nonumber \\
  & & \qquad + {1\over (2N+1)!} \left({2 \pi \over L} \right)^{2N+1}
   \int_0^{2\pi\alpha} {dp \over 2\pi} f^{(2N+1)}(p)
   \widehat{B}_{2N+1}(p)
\label{Euler_MacLaurin_formula}
\end{eqnarray}
where 
\be
\widehat{B}_n(p) = B_n\left( {L p \over 2 \pi} -
                           \left\lfloor {L p \over 2 \pi} \right\rfloor \right)
\ee
The expression \reff{Euler_MacLaurin_formula} gives the asymptotic 
expansion of the sum \reff{goal_sum} in powers of $L^{-1}$
up to order $L^{-2N}$ {\em if} the last integral in 
\reff{Euler_MacLaurin_formula} is finite (i.e., if $f^{(2N+1)}(p)$ is 
integrable in the interval $[0,2\alpha\pi]$). If $f(0) = f(2\pi\alpha)$, 
then only {\em even} powers of $L^{-1}$ occur in the expansion  
\reff{Euler_MacLaurin_formula}.
  
%
%  End text sections
%
%%%%%%%%%%%%%%%%%%%%%%%%%%%%%%%%%%%%%%%%%%%%%%%%%%%%%%%%%%%%%%%%%%%%%%%%

\section*{Acknowledgments}

We wish to thank Nickolay Izmailian for correspondence and useful 
clarifications about Ref.~\cite{Izmailian}; Birte Jackson and  
Robert Shrock for their warm hospitality and discussions during the authors' 
visit to the C.N.~Yang Institute for Theoretical Physics, where this work 
was mainly done; and Alan Sokal for useful discussions and for a critical 
reading of the first draft of this manuscript.  
The authors' research was supported in part by CICyT (Spain) grants 
AEN97-1680 and AEN99-0990.

%\newpage
%%%%%%%%%%%%%%%%%%%%%%%%%%%%%%%%%%%%%%%%%%%%%%%%%%%%%%%%%%%%%%%%%%%%%%%%
%
% Bibliography definitions
%
\renewcommand{\baselinestretch}{1}
\large

\end{document}